\documentstyle[aps,pra,twocolumn]{revtex}
\newcommand{\be}{\begin{equation}}
\newcommand{\ee}{\end{equation}}
\newcommand{\bea}{\begin{eqnarray}}
\newcommand{\eea}{\end{eqnarray}}
\newcommand{\bra}[1]{\left< #1 \right|}
\newcommand{\ket}[1]{\left| #1 \right>}
\newcommand{\proj}[1]{\left| #1 \right>\left< #1 \right|}
\newcommand{\ketbra}[2]{\left| #1 \right>\!\!\left< #2 \right|}

\newcommand{\dreiv}[3]{\left( \begin{array}{c} #1\\#2\\#3 \end{array} \right)}

\newcommand{\half}{\mbox{$\textstyle \frac{1}{2}$}}

\newcommand{\sthird}{\mbox{$\textstyle \frac{1}{\sqrt{3}}$}}
\newcommand{\stwelvth}{\mbox{$\textstyle \frac{1}{\sqrt{12}}$}}
\newcommand{\ssixth}{\mbox{$\textstyle \frac{1}{\sqrt{6}}$}}
\newcommand{\stwothird}{\mbox{$\textstyle \sqrt{\frac{2}{3}}$}}
\newcommand{\eqref}[1]{(\ref{#1})}
\newcommand{\Tr}[1]{\mbox{Tr}\left( #1 \right)}
\newtheorem{theorem}{Theorem}
\draft
\begin{document}
\title{Quantum Cloning and Distributed  Measurements}
\author{ Dagmar~Bru\ss $^1$, John~Calsamiglia$^2$ and   
                 Norbert~L\"utkenhaus$^2$\footnote{Present
affiliation: MagiQ Technologies, 275 7th Avenue, 26th floor,  New York, NY 10001, USA }}
\address{$^1$Inst. f\"{u}r Theoret. Physik, Universit\"{a}t Hannover,
                 Appelstr. 2, D-30167 Hannover, Germany\\
          $^2$  Helsinki Inst. of Physics, P.O. Box 9, FIN- 
             00014 Helsingin Yliopisto, Finland}
                
\date{Received \today}
\maketitle
\begin{abstract}
We study measurements on various subsystems of the output of a
universal $1\to 2$ cloning machine, and establish a correspondence
between these measurements at the output and effective measurements on
the original input.  We show that 
one can implement sharp effective measurement elements by measuring only
two out of the three output systems.
 Additionally, certain complete sets of sharp
measurements on the input can be realised by measurements on the two
clones.  Furthermore, we introduce a scheme that allows to
restore the original input in one of the output bits, by using
measurements and classical communication -- a protocol that resembles
teleportation.
\end{abstract}

\narrowtext

\section{Introduction}
\label{sec:intro}

The No-Cloning theorem \cite{wootters82a}
states that laws of quantum mechanics
forbid to design an apparatus which is always successful in making
 an exact copy of 
an unknown quantum state. The fact that it is possible to
 make either imperfect copies with probability one \cite{buzek96a}
or perfect copies with 
probability less than one \cite{duan98a}
is by now well established, and upper
bounds for these scenarios have been derived \cite{gisin97a,bruss98c,werner98a}.

So far this topic has been mainly addressed with the purpose to
study a fundamental concept of quantum mechanics and its 
implication in connection with quantum information. 
Neither has a cloning transformation been realised experimentally
so far,
nor has cloning been shown to be a useful concept for
quantum information processing.
Recently, though, an
experimental scheme for the realisation of an optimal  cloning
process via stimulated emission has been suggested \cite{simonsub99a}. 
Furthermore, imperfect cloning was shown to enhance the efficiency of
imperfect detectors \cite{deuarsub99a}, and to improve the
performance of some quantum computation tasks \cite{hardy00}.
 
A quantum cloning transformation spreads the 
 information that is contained in the input state over the
 entangled wave function of the output state. 
In this article we study various measurements at the
output of an optimal universal 
$1\to 2$ quantum cloning device,
i.e. a black box that takes one unknown quantum bit as
input and creates three entangled outputs. Two of those subsystems are
 the  clones, one is an auxiliary system.
 We
relate the measurements on the output systems
 to ``effective'' measurements at the input. 

When having access to the total output state, 
applying the inverse cloning
transformation evidently leads to the original input state and
thus gives access to the original information. Let us imagine
now that one has access only to a part of the output, e.g.
if one of the subsystems was lost.
Can one still make a measurement at the output that corresponds to a sharp 
measurement at the input?
Our counterintuitive result is that this is indeed the case, when
one has access to the two clones only. 
It turns out that the sharp measurements which can be implemented 
in this
way are capable to extract the {\em maximum} accessible Shannon information
on the -- uniformly di tributed -- 
input state. On the other hand, by measurements 
on the ancilla one can also 
access {\em some} Shannon information on the input state. (The
quantitative calculation is an adaption of that given in \cite{peres93a}.) 
In this  sense the information contained in the input is spread in a partly 
redundant way over the output of the cloner.
This property might be of use in potential 
quantum error correction schemes
based on quantum cloning.

This paper aims at a systematic study of the distribution of information
at the cloning output by investigating general measurements of various 
combinations of subsystems
of the output. 
Questions regarding the mutual information for one-particle
subsystems only at the cloning output have  been addressed in \cite{munro}.
Here we also study the effect of the entanglement at the output
and derive  explicit expressions for effective measurements
 on the input that correspond to
  measurements performed on subsystems of
the output.

A further motivation for our study is related to the present search for
better understanding of multiparticle entanglement: the cloning
output is a  special 
three-party entangled state. 
We will show that its
peculiar properties allow to restore the original state in one of
the output qubits by performing 
certain measurements on the other qubits
and communicating classically -- a protocol
that bears a likeness to teleportation.

Note that one cannot find an analogous state for a two qubit 
system by
 studying the output of a universal $1\to 1$ cloner
\cite{bruss99a}: the fidelity $F=1$ is reached by applying the identity,
and achieving a fidelity smaller than one requires
an ancilla with a dimension higher than two. 

Our work
is organised as follows: in section \ref{sec:clone} we remind the
reader of the optimal $1\to 2$ cloning transformation.
 In section \ref{sec:measure}
we present measurements 
of various subsystems of the cloning output
and compare with the according effective measurements at the input.
Thus we 
show that it is enough to possess two 
 subsystems of the output
in order to perform measurements that correspond to sharp 
measurements at the input.
In section \ref{sec:comm} we introduce the possibility
of restoring the input in a subsystem of the output state by 
performing a  
measurement  on a different output subsystem
and then  communicating classically.
Finally, we give a summary.
 
 \section{Revisiting the universal $1 \to 2$  cloner}
\label{sec:clone}
The universal $1 \to 2$  cloner in two dimensions
takes as input a
general pure input state 
$\ket{\varphi}=\alpha \ket{0}+\beta \ket{1}$,
the density matrix of which will be written as 
\begin{equation}
\proj{\varphi}=
\rho_{in} = \frac{1}{2} \left( \openone + \vec{s}_{in} \cdot \vec{\sigma} \right)
\label{defs}
\end{equation}
 where $ \vec{\sigma} = \{\sigma_x,\sigma_y,\sigma_z\}$ are the Pauli
matrices and 
$ \vec{s}_{in} = \{s_x,s_y,s_z\}$ defines the input Bloch vector
 with unit length.

A cloning transformation $U$ 
is specified through  its action on the basis
states.
The family of optimal universal $1 \to 2$ 
cloning transformations  is
given in
 \cite{bruss98d}. 
 A special choice of
 phases and the ancilla  leads to the transformation which we
 will use throughout this paper, namely 
\begin{eqnarray}
\label{1to2cloner}
U \ket{0} \ket{0} \ket{0} & = & \sqrt{\frac{2}{3}} \ket{00}
\ket{1}  - \frac{1}{\sqrt{6}} (\ket{01}+\ket{10})
\ket{0} \  ,\\
 U \ket{1} \ket{0} \ket{0} & = &  -\sqrt{\frac{2}{3}} \ket{11}
\ket{0}  + \frac{1}{\sqrt{6}} (\ket{10}+\ket{01})
\ket{1} \  ,
\end{eqnarray}
where the first and second bit on the right hand side
refer to the clones, and the third bit is the ancilla.
Note that this transformation differs in the ancilla from
the one considered in \cite{buzek96a}. Our choice is such that the
Bloch vector of the ancilla's reduced density matrix 
 is given by $\vec{s}^{a}=-{1\over3} \vec{s}_{in}$. This results
in a special symmetry of the cloning transformation which we will
introduce later.

We will recall further
properties of this transformation in the next sections,
wherever they are needed. 

The connection between optimal 
quantum cloning and optimal state estimation has been 
established in \cite{bruss98c} for the case of infinitely many copies.
In our paper we are interested 
in  measurements at the output of an optimal  
cloner for the case of two
copies only.

\section{Measurements on subsystems of the output}
\label{sec:measure}
\subsection{Effective POVM's}
\label{effpom}
The aim of 
 this section is to  relate possible measurements on the input
 $\rho_{in}$ 
to measurements on parts of the output of the cloner, namely on 
one clone, or the ancilla, or the ancilla and one clone, or
on two clones. 
For this purpose we will introduce the notion of {\em effective} POVM's.

A general measurement or \emph{positive operator valued measure} 
(POVM) \cite{peres93a}
on a quantum state $\rho$ is described by a set 
of operators $\{ F_{i} \}$
 which is a resolution of the identity,
\begin{equation}
	\sum_{i}^{n}F_{i}=\openone\ ,
	\label{eq:res.identity}
\end{equation}
where each \emph{POVM element} $F_{i}$ is a positive operator and is
associated with one measurement outcome such that the probability of
occurrence of this outcome is given by $p_{i}=\mbox{Tr}(F_{i}\rho )$.

The cloning transformation connects  the input  and the
output state unitarily.  
Two of the input qubits 
(blank and auxiliary qubit) are in a known prescribed state.
Any measurement defined by a POVM on the output system $\rho_{out}$
can be alternatively described by an {\em effective} POVM  on the input
state $\rho_{in}$,  using the  equality
\begin{eqnarray}
\mbox{Tr}[F_i \rho_{out}]  &=&  \mbox{Tr}[F_i U 
  \rho_{in}\otimes\ket{00}_{ba}\bra{00} U^{\dagger}] = \nonumber\\ 
    &=&  Tr[_{ba}\!\bra{00}U^{\dagger} F_i U \ket{00}_{ba} \rho_{in}]
    \nonumber \\ 
	 &=&\mbox{Tr}[E_i \rho_{in}] 
	 \label{eq:totalPOVM}
\end{eqnarray}
where
$\ket{00}_{ba}$ denotes the blank  and the  ancilla input.
Thus we have found for any
 POVM element $F_i$ on the output Hilbert space
the  corresponding {\em effective POVM element}  $E_i$   
 on the input as
\begin{equation}
	E_i= \, _{ba}\!\bra{00}U^{\dagger} F_i U \ket{00}_{ba}\mbox{.} 
	\label{eq:Efpovm}
\end{equation}

In the following subsections we proceed to calculate the effective
POVM elements $E_i$ for  measurements which act 
on certain subsystems of the output, for example POVM
elements $F_i^c \otimes \openone^c \otimes \openone^a$ which are
measurements on just one clone. 

Which are the effective measurements on the input that correspond to
``valid'' measurements on the output? These effective POVM's  have to satisfy 
two criteria:
(a) each element of the effective POVM must be realizable by a
corresponding POVM element on the  output  and (b) the
collection of corresponding POVM elements at the output 
must be complete, i.e. they have to add up to the identity.
As we will see, we can
 establish this correspondence for
 several  POVM's on certain subsystems
of the output. 
 
Our main interest lies in sharp POVM elements, that is
operators of rank one. The occurrence of the result of a sharp measurement 
allows to exclude states orthogonal to the support of
its corresponding matrix. Complete POVM's formed by sharp measurement
elements play an important role in extracting information
about the input state.
 It was already shown by Davies \cite{davies78a} that 
one can always optimize the accessible Shannon information by a POVM
with sharp elements. If the set of input states has uniform
distribution of the Bloch vector over the unit sphere, then 
 actually {\em any} POVM formed only by 
sharp elements optimizes the accessible Shannon information. This
is due to  isotropy, since in this case the occurrence of each
measurement outcome reveals the same amount of information about the
input state. This statement
can also be checked explicitly by calculations following
those given in \cite{peres93a}. Moreover, optimal state estimation for
isotropically distributed one-qubit states \cite{massar95} can be realised
by any sharp POVM. This can be seen from work by Vidal et al.~\cite{vidal99a}.

\subsection{General properties of measurements on subsystems} 
\label{genprop}

In the following, we will investigate in which situation it is
possible to perform a sharp measurement on the input system by
measuring only a subsystem of the output of the $1\to 2$ cloner, for
example by measuring the two clones, but not the auxiliary system. In
this investigation  the following theorem is useful, which states
that for each output POVM element belonging to this class
 the conditional density matrix of the unmeasured system must
necessarily be in a state independent of the input state.

\begin{theorem} 
\label{sharpnesstheorem}
Consider a product Hilbert space of system A and B prepared in a
product state described by a density matrix $\rho_A \otimes
\rho_B$. Let $U$ be a  unitary operator that
maps the Hilbert space ${\cal H}_A \otimes {\cal H}_B$ onto a
 Hilbert space of the same dimension with a product structure
${\cal H}_C \otimes {\cal H}_D$. 
Given the POVM element $F$ that acts on 
system $C$,  the corresponding 
effective POVM element $E$ on system
$A$  is described by
\begin{equation}
E = \mbox{Tr}_B \left( \rho_B U^\dagger F \otimes \openone_D U \right) \; .
\end{equation}
If the operator $E$ on ${\cal H}_A$ is of rank one (sharp POVM element), then 
the
state of system $D$, conditioned on  the measurement
outcome at $C$,  is independent of
$\rho_A$.  
\end{theorem}

The statement of the theorem can be rephrased by viewing this procedure as a quantum
channel which maps the input state $\rho_A$ 
onto  the outgoing subsystem 
$\rho_D=\mbox{Tr}_C\left(U \rho_A \otimes \rho_B U^\dagger \right)$ which is
{\em not} measured. This mapping is  completely positive (CP), and
 can be written in terms of the Kraus operators $\{A_i\}$, 
\begin{equation}
\rho_A \to \rho_D=  \sum_{i=1}^{N} A_i\rho_A A_i^\dagger \; ,
\end{equation}
where $\sum_{i=1}^{N} A_i^\dagger A_i=\openone$.
If we consider  only one measurement outcome, as in the theorem, then
the {\em conditional} state corresponding to this outcome is given by
\begin{equation}
 \rho_D^{cond}= \frac{1}{M}\sum_{i \in K} A_i \rho_A A_i^\dagger \; ,
\end{equation}
where $M$ is a
normalization constant, and
 $K$ is some subset of the index set $i=\{1, \dots N\}$. In the most
simple case it will contain only one element.  

The effective POVM element 
corresponding to  this measurement outcome is therefore given by
the Kraus operators as 
\begin{equation}
\label{KrausPOVM}
E=\sum_{i\in K} A_i^\dagger A_i \; .
\end{equation}
If the operator $E$ is of rank 1, it can be written as
 $E=p
\ket{\tilde{\chi}}\bra{\tilde{\chi}}$ with some $p$ satisfying $0 \leq
p \leq 1$. Since any operator $A^\dagger_i A_i$ is  positive,
 the structure of Eqn.~(\ref{KrausPOVM}) implies that
 each operator $A^\dagger_i A_i$ is of rank 1. We make a singular
 value decomposition ansatz  $A_i=V S \tilde{V}$, where $S$ is a
 positive, real diagonal matrix, and $V, \tilde{V}$ are unitary
 matrices. It follows that $S$ must be of rank 1.  
Therefore, we can write each Kraus operator as $A_i
= \gamma_i
\ket{\chi_i}\bra{\tilde{\chi}}$ with $\gamma_i$ real and positive and
$\sum\gamma_i^2
=p$, where the vectors
$\ket{\chi_i}$ might be different for the various contributing
Kraus operators, while the vector $\ket{\tilde{\chi}}$ is fixed by
$E$. This immediately gives us the final state
\begin{eqnarray}
\rho_D^{cond} &=&\frac{1}{M}
\bra{\tilde{\chi}}\rho_A\ket{\tilde{\chi}} \sum_{i
\in K}\gamma_i^2 \ket{\chi_i}\bra{\chi_i}\\
& =& \frac{1}{M'} \sum_{i
\in K}\gamma_i^2 \ket{\chi_i}\bra{\chi_i}\; .
\end{eqnarray}
Here $M$ and $M'$ are normalization factors. The final conditional state does no
longer depend on the input state. This proves our theorem.

\subsection{Measurements on one clone or on the ancilla}

The most general POVM element on one single qubit 
of the output is given by
the expansion of a hermitian matrix in terms of the Pauli matrices,
namely 
\begin{equation}
F_i = b_i \left(\openone + \vec{f}_i \cdot \vec{\sigma} \right)
\label{eq:povm1cl}
\end{equation}
where $0<b_i\leq1$ and $|\vec{f}_i|\leq 1$ for positivity.
(A sharp POVM element is characterized by  $|\vec{f}_i|= 1$.)
The corresponding effective POVM element 
at the input can be written in the same form as 
\begin{equation}
E_{i} = a_i \left(\openone + \vec{e}_i\cdot \vec{\sigma} \right)\ .
\label{eq:inPOVMec}
\end{equation}
In order to find the parameters $a_i$ and $\vec{e}_i$ as functions of
$b_i$ and $\vec{f}_i$
we use the equality
\begin{equation} 
\mbox{Tr}(E_{i} \rho_{in})
= \mbox{Tr}( F_i \rho_{out,red})\; ,
	 \label{eq:eq1cl}
\end{equation}
where $\rho_{out,red}$ refers to the according subsystem at the
output. 
We find immediately
\begin{equation}
	a_i \left(1+ \vec{s}_{in}\cdot\vec{e}_i \right)=b_i (1+ 
	\vec{s}_{out,red}\cdot\vec{f}^c_i ) \; .
	\label{eq:prbanc}
\end{equation}
Here $\vec{s}_{out,red}$ denotes the Bloch vector of the output state.
For the universal cloner the Bloch vectors of both clones have the
same orientation as for the input \cite{bruss98d}. The Bloch vector of
the ancilla can be chosen to be antiparallel to that \cite{buzek99a}.
 After applying
the optimal transformation the Bloch vector for a clone 
is shrunk by a factor of 2/3
with respect to the input, and 
we denote it as $\vec{s}^c=2/3\,\cdot  \vec{s}_{in}$.
The Bloch vector  of the ancilla is here given by
$\vec{s}^a=-1/3\, \cdot \vec{s}_{in}$, according to the
 transformation $U$  in equation (\ref{1to2cloner}).

Since equation (\ref{eq:prbanc})
has to  hold for  all input states and therefore for all Bloch
vectors $\vec{s}_{in}$, we find for the parameters 
of measurements on one clone, denoted by the superscript $c$:
\begin{eqnarray}
a^{c}_i& = & b^{c}_i\ ,\\
\vec{e}^c_i & = & \frac{2}{3} \vec{f}^c_i
\end{eqnarray}

Thus only effective POVM elements with $|\vec{e}^{c}_i|
 \leq 2/3$ can be realized by a measurement on one clone alone;
  sharp measurements like projection measurements are excluded.
 This is consistent with the fact that the clone can be understood 
 as a mixture of a completely random state with probability $\frac{1}{3}$ 
 and the input state with probability $\frac{2}{3}$.
So,  the result of passing our original state through  
a Stern-Gerlach apparatus which fails with  probability $\frac{1}{3}$ would be 
 the same as the one obtained by running one clone 
 through a perfect Stern-Gerlach apparatus.

An analogous result is obtained for measurements on the
ancilla. We label the parameters of the POVM 
element in equation (\ref{eq:povm1cl}) 
with a superscript $a$ for this case and find
\begin{eqnarray}
a^{a}_i& = & b^{a}_i\ ,\\
\vec{e}^{a}_i & = & -\frac{1}{3} \vec{f}^a_i \; .
\end{eqnarray}
As expected,  measuring the ancilla alone
 cannot correspond to making sharp measurements on the
input.

We infer that the prescription of a complete POVM
on the input system defines a unique and valid POVM on one clone 
(or on the
ancilla), provided that for each 
POVM element the relation $|\vec{e}^{c}_i|
\leq 2/3$ (or $|\vec{e}_i^a| \leq \case{1}{3}$) holds. The completeness
relation for measurements at the input reads
$\sum_i a_i=1$ and $\sum_i a_i \vec{e_i}=0$. If these
equations  hold,  
the completeness
relation for measurements at the output, namely
$\sum_i b_i=1$ and $\sum_i b_i \vec{f_i}=0$,
is here automatically satisfied,
due to the equality $a_i=b_i$ and the 
fact that $\vec{e}_i$ and $\vec{f}_i$ are related by a constant
factor.

\subsection{Coherent measurements on two subsystems}
\label{sec:twoclones}
In the previous section we investigated, which effective POVMs can be
realized by measurements on single subsystems of the output of the
$1\to 2$ cloner. In this section we  consider coherent
measurements on two subsystems.

Before studying these specific measurements, let us first
introduce general measurements on higher-dimensional spaces:
 any  POVM element acting in a $d$-dimensional Hilbert space
 can be written in the form
\begin{equation}
F_i = b_i \left(\openone + \vec{f}_i\cdot\vec{\tau} \right)
\label{eq:pov}
\end{equation}
where the parameters
$b_i$ and $\vec{f}_i$ are real,
and $\vec{\tau}$ denotes the $d^2-1$ generators of  SU($d$),
which obey Tr$(\tau_i\tau_j)=2\delta_{ij}$.
Which are the
 necessary  conditions, such that equation (\ref{eq:pov}) describes 
 a valid POVM element? 
We will only give two obvious conditions. 
A complete and explicit characterisation of necessary {\em and}
sufficient conditions is beyond the scope of this paper.
As the eigenvalues of each POVM element have to be in the
  interval $[0,1]$, the dimension $d$ of the underlying Hilbert space
  leads to the constraint  $0\leq \Tr{F_i}\leq d$, 
and therefore  to $0 \leq b_i \leq 1 $. 

Another necessary condition is $|\vec{f}_i|^2 \leq d(d-1)/2$, for the following
reason:  a projection measurement
 can be written in the form $F_i=\ket{\phi}\bra{\phi} =
\frac{1}{d}\left(\openone + \vec{f}_i\cdot\vec{\tau} \right)$, and the
condition $\Tr{F^2_i}=1$ then leads  to $|\vec{f}_i|^2=d(d-1)/2$.
 Since a general POVM element is proportional to a convex combination of
   projectors,  we obtain in general  $|\vec{f}_i|^2 \leq d(d-1)/2$. 
 
\subsubsection{Measurements on the two clones}

Let us turn to measurements on the two clones.
For our purposes
it is convenient to rewrite the cloning transformation (\ref{1to2cloner}),
using the Bell basis for the two clones: 
\begin{eqnarray}
\label{cloner}
U \ket{0} \ket{0} \ket{0} & = & \frac{1}{\sqrt{3}} \ket{\phi^+}
\ket{1} + \frac{1}{\sqrt{3}} \ket{\phi^-}
\ket{1} - \frac{1}{\sqrt{3}} \ket{\psi^+}
\ket{0} \  ,\\
 U \ket{1} \ket{0} \ket{0} & = &  -\frac{1}{\sqrt{3}}\ket{\phi^+}
\ket{0} +  \frac{1}{\sqrt{3}}\ket{\phi^-}
\ket{0} + \frac{1}{\sqrt{3}} \ket{\psi^+}
\ket{1} \  ,
\end{eqnarray}
where the Bell states are defined as
\begin{eqnarray}
\ket{\phi^{\pm}} & = & 1/\sqrt{2} \left( \ket{00}\pm\ket{11} \right) \  ,\\
\ket{\psi^{\pm}} & = & 1/\sqrt{2} \left( \ket{01}\pm\ket{10} \right) \ .
\end{eqnarray}
Only the symmetric Bell states appear in the cloning 
transformation, because the two clones are
required to be in the symmetric subspace \cite{werner98a}.
Their reduced
density matrix can  be written in a simple form
 in the Bell basis $\{ \phi^+,\psi^+,\phi^-, \psi^- \}$: 
\begin{equation}
		 \rho^{cc}=
  {\mbox{Tr}_{a}}\varrho_{out} =
  \frac{1}{3} \left(
  \begin{array}{llll}
              1 & s_x & s_z & 0 \\
              s_x & 1 & i s_y & 0 \\
              s_z & -i s_y & 1 &0 \\
   0 & 0 & 0 & 0
            \end{array} 
			\right) \; 
			\label{eq:rhocc}
	\end{equation}
where the trace is performed over the ancilla. 

As this  density matrix has only support in the 
3-dimensional symmetric subspace, we can decompose it
using the eight generators of SU(3),
namely   $\lambda_k$ (which are labelled
in a standard way, see e.g. \cite{greiner94a}),
 \begin{equation}
 \rho^{cc} =  \frac{1}{3} \left( \openone + \dreiv{s_x}{-s_y}{s_z}
	\dreiv{\lambda_1}{\lambda_7}{\lambda_4} \right) = \frac{1}{3} \left( 
  \openone+\vec{s}^{cc} \vec{\lambda}  \right) \ \ .
\label{eq:redcc}
\end{equation}
 
 Note that the eight-dimensional 
 generalized Bloch vector $\vec{s}^{cc}$ has only three
 non-vanishing entries.

For measurements on the two clones it is sufficient to consider 
POVM's on the symmetric subspace of two qubits.
 Any such POVM element can be written in the form
\begin{equation}
F^{cc}_i = b^{cc}_i \left(\openone + \vec{f}^{cc}_i\cdot\vec{\lambda} \right)
\label{eq:povcc}
\end{equation}
where the parameters
$b^{cc}_i$ and $\vec{f}^{cc}_i$ are real, with the constraints
$0 \leq b^{cc}_i \leq 1 $, and 
 $|\vec{f}^{cc}_i|^2 \leq 3$, as explained below equation (\ref{eq:pov}). 
  Any anti-symmetric
component of a measurement on the two clones has no effect on the
effective POVM.  
 
 As in the previous subsection we can now relate the parameters of
 an effective POVM element on the
input Hilbert space,
 \begin{equation}
	E^{cc} = a^{cc}_i \left(\openone + \vec{e}^{ cc}_i
	\cdot\vec{\sigma} \right)\ ,
\end{equation}
to the parameters in equation (\ref{eq:povcc}) via the
equality
\be
Tr[E^{cc}_i\rho_{in}  ]=Tr[F^{cc}_i \rho^{cc} ] \ ,
\ee
which leads to the assignment
\begin{eqnarray}
a^{cc}_i & = & b^{cc}_i \ , \nonumber \\
\vec{e}^{cc}_i & = &
\frac{2}{3}\dreiv{f^{cc}_{1}}{-f^{cc}_{7}}{f^{cc}_{4}}_i \; .
\label{eq:solcc}
\end{eqnarray}
Note that one can show that this relation can be expressed more simple as
\begin{equation}
\label{EFformula}
E_i^{cc} = \frac{2}{3}  Tr_c \left( F^{cc}_i \right)
\end{equation}
where the trace is performed over one of the clones and the resulting
operator is interpreted as acting on the input state. 

Given the output POVM element $F_i$, this assignment uniquely determines the
effective POVM element $E_i$ on the input. However, we see that conversely the
prescription of the effective POVM $E_i$ leaves in general five parameters
of the output POVM element $F_i$ undetermined. 
In order to establish whether a
particular effective POVM element can be realised by a measurement on
the two clones, one  needs to show that there exists a choice of
the undetermined parameters $f^{cc}_{\nu}$ with $\nu=2,3,5,6,8$, such that
the resulting operators $F^{cc}_\nu$ are valid POVM elements. 

Here a complete POVM on the input does not necessarily
correspond   to
a complete POVM at the output. In some cases, though, 
choosing the free parameters for the output
POVM elements appropriately will allow 
the output POVM 
to fulfill the completeness relation.

We will now derive a complete set 
of sharp measurements on the two clones which 
  corresponds to rank 1 effective POVM
elements on the input. The derivation uses a symmetry argument and
the general statement about
sharp measurements given in section \ref{genprop}.

The optimal universal $1\to 2$ cloning transformation $U$
obeys the following symmetry  for
 any matrix $V \in SU(2)$:
\begin{equation}
U \left(V \otimes \openone \otimes \openone \right) \ket{000} = 
\left(V \otimes V \otimes V\right) U \ket{000} \; .
\label{vsymmetry}
\end{equation}

According to section \ref{genprop} 
the sharp measurement on the two clones has to leave  the ancilla
in a state independent of the input.
 The symmetry property  (\ref{vsymmetry}) allows us
to consider a fixed state of the ancilla, and we choose the
 state $\ket{0}$.
 
 Given a general input state $\ket{\varphi}$, 
 inspection of the cloning transformation (\ref{1to2cloner}) 
   leads to  the 
  only choice for a symmetric 
  projection measurement on the clones that leaves the
  ancilla in the state  $\ket{0}$, namely
  $F_i^{cc} = p\proj{11}$ with $0\leq p \leq 1$.
  Using equation (\ref{EFformula}) 
  it is straightforward to calculate the corresponding effective POVM
 element, which is given by $E^{cc}_i= \frac{2}{3} p
\ketbra{1}{1}$. 
 
 Making use of the symmetry of the universal cloner we find  
that 
the corresponding complete class of  POVM elements on the symmetric
subspace of the two clones leading to sharp effective POVM elements on
the input is given by
\begin{equation}
\label{sharp2clones}
F_i^{cc} =
 p\ketbra{\chi_i}{\chi_i}\otimes \ketbra{\chi_i}{\chi_i} \otimes 
 \openone_a \ .
\end{equation}
The corresponding  sharp effective POVM elements on the input are 
\begin{equation}
E^{cc}_i = \frac{2}{3} p \ketbra{\chi_i}{\chi_i} 
\end{equation}

Clearly, we cannot implement a measurement like in a Stern Gerlach apparatus, 
since this would
correspond to $p = 3/2$.  However, one can e.g. 
choose a mixture of three set-ups, with equal probability, discriminating in
the $x,y$, and $z$-direction. 
The complete set of effective POVM
elements is then  given by 
\begin{equation}
\{E_i^{cc}\} =
\{\frac{1}{3}P_{x+},\frac{1}{3}P_{x-},\frac{1}{3}P_{y+},\frac{1}{3}P_{y-},
\frac{1}{3}P_{z+},\frac{1}{3}P_{z-}\}\ ,
\end{equation}
 which   corresponds to the  complete set 
 of POVM elements on the two clones 
\begin{eqnarray}
\label{complete2clones1}
\{F_i^{cc}\} &=&
\{\frac{1}{2} \ket{x+}\ket{x+}\bra{x+}\bra{x+},\frac{1}{2}
\ket{x-}\ket{x-}\bra{x-}\bra{x-}, \nonumber \\
 &&  \; \frac{1}{2} \ket{y+}\ket{y+}\bra{y+}\bra{y+},\frac{1}{2}
\ket{y-}\ket{y-}\bra{y-}\bra{y-}, \nonumber \\
 && \; \frac{1}{2}
   \ket{z+}\ket{z+}\bra{z+}\bra{z+},\frac{1}{2}
\ket{z-}\ket{z-}\bra{z-}\bra{z-}\}\ .
\end{eqnarray}

Another possibility is given by the effective POVM
\be
\{E_i^{cc}\} =
\{\frac{1}{2}\proj{\vec{n}_1},\frac{1}{2}\proj{\vec{n}_2},
\frac{1}{2}\proj{\vec{n}_3},\frac{1}{2}\proj{\vec{n}_4}\}\ ,
\ee
where the Bloch vectors $\vec{n}_i$ point to the corners of a
regular tetrahedron.
This corresponds to measurements at the output
\begin{equation}
\label{complete2clones2}
\{F_i^{cc}\} =
\{\frac{3}{4}\proj{\vec{n}_1\vec{n}_1},\ldots\frac{3}{4}\proj{\vec{n}_4
\vec{n}_4}\}\ .
\end{equation}
This measurement corresponds to  optimal
state estimation \cite{massar95}  on two identical qubits that
have uniform a-priori
probability distribution. 

Note that
 the POVM elements (\ref{sharp2clones}) are products of POVM
elements on each individual clone. Therefore, it is possible to
implement by means of local operations and classical communication
(LOCC) a POVM on the two clones which contains {\em some} sharp
effective POVM elements.
However, it can be seen that a complete set of such elements as in 
(\ref{complete2clones1}) or (\ref{complete2clones2}) cannot be
realised by LOCC measurements. The reason for this is that any
complete POVM realised by LOCC operations  has an antisymmetric
component. Extending or completing our POVMs (which were restricted to
the symmetric space for convenience) will always lead to non-local features. 

 In this section we have found the surprising result that 
sharp effective measurements on the input can be achieved when 
measuring the two clones only, i.e. the information that is
contained in the ancilla is in this sense redundant.
 The examples of effective
POVM's we presented  are actually suited to optimise 
the fidelity of state estimation
or the accessible Shannon information for a uniformly distributed input
qubit. These tasks can therefore be performed equally well with the
input state or the two clones alone, see \cite{bruss98c} and
 \cite{vidal99a}.


\subsubsection{Measurements on one clone and the ancilla}

The reduced density matrix of the two subsystems formed by  one clone 
and one ancilla  acquires the following form if expressed in the Bell basis:

\begin{equation}
\rho^{ca}=\frac{1}{12}\left(
 \begin{array}{cccc}
	 1& s_{x} & s_{z} &  3 i  s_{y}\\
	s_{x} &  1& i s_{y} & 3 s_{z} \\
	s_{z} & -i s_{y} & 1 & -3 s_{x}  \\
	 -3 i s_{y} & 3 s_{z} & -3 s_{x} & 9
  \end{array}\right)\ ,
	\label{eq:redac}
\end{equation}
which does have support on the 
antisymmetric space, i.e. the entries for $\ket{\psi^-}$
(last row and column) do not vanish.

However, comparing  (\ref{eq:redac}) with 
 the density matrices of the two
clones (\ref{eq:rhocc}), 
 we note that the symmetric part for the clone-ancilla system is
equal to the density matrix of the two clones times the factor
$\case{1}{4}$. Consequently, all sharp effective POVM elements $E_i$ 
corresponding to measurements $F_i$ on
the clone-clone system can be
realised by the same measurements $F_i$ on the clone-ancilla system, 
although the weight $a_i$ in the effective POVM $E_i$ 
will be decreased by a
factor $1/4$. 

As in the previous paragraph about measurements on two clones one can
also derive a correspondence like in  (\ref{eq:solcc}), where
the vector $\vec f$  will
now be 15-dimensional, as the general POVM has to be expanded in terms
of the SU(4) generators \cite{greiner94a}. We give this correspondence
for completeness:
\begin{eqnarray}
a^{ca}_i & = & b^{ca}_i(1-\sqrt{\frac{2}{3}}f^{ca}_{15})\ , \nonumber   \\
\vec{e}^{ca}_i & = & \frac{1}{6 (1-\sqrt{\frac{2}{3}}f^{ca}_{15})}
\dreiv{f^{ca}_{1}-3 f^{ca}_{13}}{-f^{ca}_{7}-3f^{ca}_{10}}
{f^{ca}_{4}+3f^{ca}_{11}}_i \; .
\label{eq:solca}
\end{eqnarray}

Can we  find a complete   measurement on a 
the clone-ancilla subsystem
of the output that corresponds to sharp measurements at the input,
like in the previous subsection?

 We follow the same arguments as in the case of measurements on
the two clones.
  First, we rewrite 
  the cloning transformation (\ref{1to2cloner}), now
   sorting the terms according
to the value of the first qubit: 
\begin{eqnarray}
U \ket{000} & = & \frac{1}{\sqrt{12}}\left( \ket{0} \left(\ket{\psi^+} + 3
\ket{\psi^-}\right) - \ket{1}\left( \ket{\phi^+}+\ket{\phi^-}\right)
\right)\ , \\
U \ket{100} & = & \frac{1}{\sqrt{12}}\left( -\ket{1} \left(\ket{\psi^+} -3
\ket{\psi^-}\right) + \ket{0}\left( \ket{\phi^+}-\ket{\phi^-}\right)
\right) \ .
\end{eqnarray}
By demanding (compare the argument in  the previous section)
that the first clone should be in state $\ket{0}$
after a projection measurement, we find
that the  measurement 
$F_i = \proj{\kappa_{F_i}}$ on the other clone and the ancilla 
has to be given by
\begin{equation}
\ket{\kappa_{F_i}}=A \left( 3 \ket{\psi^+} + \ket{\psi^-} \right) + B \left(
\ket{\phi^+} -
\ket{\phi^-}\right)
\label{kappa}
\end{equation}
where  $A$ and $B$ are free parameters.

Using (\ref{eq:Efpovm})
it is straightforward to check that the corresponding effective POVM
is indeed of rank one,
i.e. $E_i=\proj{\kappa_{E_i}}$,
 and is given by
\begin{equation}
\label{kappastates}
\ket{\kappa_{E_i}}=\frac{1}{\sqrt{12}} \left( 6 A \ket{0} + 2 B \ket{1} \right) \; .
\end{equation}

Can we find a complete POVM at the output, i.e.
a set $\{F_i\}$ with
$\sum_iF_i=\openone$,  that corresponds 
to sharp POVM elements on the input?
Given the symmetry of the cloning transformation 
we know that all the POVM elements $F_i$ on one clone and ancilla
  must be projectors onto states of the
form $V \otimes V \ket{\kappa_{F_i}}$,  with $ \ket{\kappa_{F_i}}$
given in equation
(\ref{kappa}).
The antisymmetric subspace is invariant under the unitary
transformation $V \otimes V$, and 
a symmetric state remains symmetric after applying $V \otimes V$.
Therefore the weight of the symmetric subspace in {\em any} projector 
onto $V \otimes V \ket{\kappa_{F_i}}$ is at least 9 times as
much as the weight of the antisymmetric subspace,
due to the term proportional to $A$ in (\ref{kappa}). A sum
of such projectors cannot be a resolution of the identity,
as the weight of the symmetric subspace in the identity is only
three times as much as for the antisymmetric subspace.

Thus we cannot find a complete POVM acting on one
clone and the ancilla  that corresponds 
to sharp POVM elements on the input, despite the fact that we can realize
individual sharp POVM elements. Actually, some of these individual
sharp effective POVM elements can be realised by local (LOCC) means
since the states (\ref{kappa}) contain product states, for
example for $A=0$. 

\section{Restoration via local measurements and communication}
\label{sec:comm}
 In the previous section we investigated which
measurements are possible if one has access only to some subsystem of the
output of the $1\rightarrow 2$ universal cloner, instead of having
access to the whole output or to the original input state. In this
section we will study the situation in which 
the output subsystems are distributed to several parties to whom we
refer as Alice, Bob and, in some situations, Charlie.
The parties are allowed
to communicate classically with each other.

Actually, the situation with respect to effective POVM elements is
very
simple in this scenario where we have access to the complete output.
 Trivially, we can implement a complete set of effective sharp POVM's. This
follows from the fact that {\em any} projection 
of the complete output onto a state
$\ket{\Psi_{out}}$  
results in a sharp effective measurement on the
input system: 
following the formalism given in
Eqns.~(\ref{eq:totalPOVM}) and (\ref{eq:Efpovm}) we find the effective
POVM to be a projection onto the (unnormalized) state
$\bra{00}U^\dagger\ket{\Psi_{out}}$ of the input system. Here $\bra{00}$
refers to the intitial state of blank and ancilla qubit. Any sharp POVM
 on the complete output system
therefore leads immediately to a sharp effective POVM on the input system.

The interesting result shown in this part of the paper is that it is
possible to recover the original state in one subsystem by a process
resembling quantum teleportation \cite{bennett93a}. Naturally,
whenever restoration is possible, any effective POVM can be trivially
implemented by applying the corresponding POVM to the restored
qubit. 

We will discuss different scenarios: 
Perfect restoration of the input 
is possible if one party holds two subsystems and a second party
holds one subsystem, in which the original will be restored.
Probabilistic restoration in the ancilla can be achieved when 
each subsystem is held by a different party.

To begin with, let us describe a general restoration process
of the unknown input qubit in one subsystem of the output. In the
first step, the subsystems, except the target system, are being measured
and the measurement results are transmitted to the site of the target
qubit. In the second step, the owner of the target system acts on it,
for example by executing an unitary operation or by performing a
 generalized measurement. 
 The desired  result is  that the
target bit will be in the unknown state of the input bit. If this is
possible for all measurement outcomes in both steps, then this process
is deterministic, otherwise we speak of probabilistic restoration.

Both steps can be described by completely positive maps. The first step,
described by Kraus operators $A_i$, map the input system onto the
target qubit. The second step is described by Kraus operators
$B_j^{(i)}$ which depend on the measurement result of the first step
and map the state of the target qubit onto itself. In its full
generality, the resulting density matrix of the target qubit,
conditioned on measurement results in both steps, is given
by
\begin{equation}
\tilde{\rho}_{out}^{cond} = \sum_{j \in L(K)} \sum_{i \in K} B_j^{(i)}
A_i \rho_{in} A_i^\dagger {B_j^{(i)}}^\dagger \; .
\end{equation}
The index sets $K$ and $L(K)$ arise since each measurement outcome in
the two steps could be described in full generality by more than one Kraus
operator. Each event which successfully restores the input state in
the target qubit must satisfy the equation (for some positive $p$)
\begin{equation}
 \sum_{j \in L(K)} \sum_{i \in K} B_j^{(i)}
A_i \rho_{in} A_i^\dagger {B_j^{(i)}}^\dagger = p \rho_{in}
\label{pro}
\end{equation}
for all states $\rho_{in}$, especially for pure states.
 
Pure states, however, are extreme points of the convex set of all states and can
therefore not be written as convex sums of different density
matrices. Consequently, we find the stronger condition
 \begin{equation}
\label{singlecond}
 B_j^{(i)} A_i \rho_{in} A_i^\dagger {B_j^{(i)}}^\dagger = p_{ij} \rho_{in}
 \label{prop}
\end{equation}
which holds for all $j \in L(K)$ and $i \in K$ with some nonnegative
$p_{ij}$ for all pure $\rho_{in}$. 
Due to linearity, equation (\ref{prop}) also has to hold for all mixed states 
$\rho_{in}$. 
Therefore we arrive at the condition $B_j^{(i)}A_i\propto \openone $
        for  $p_{ij}\neq 0$, and   $B_j^{(i)}A_i=0$ for  $p_{ij}= 0$.
In conclusion, we
find that 
\begin{equation}
 \sum_{j \in L(K)} \sum_{i \in K}A_i^\dagger
{B_j^{(i)}}^\dagger B_j^{(i)} A_i = p \openone \; .
\end{equation}
This means that, as a principle, the event of a successful restoration
does not leak any kind of information about the restored state to the
restoring parties.
This statement cannot be reversed in general, but it can be used as a
guidance to find procedures for restoration.

Following these general considerations, we can distinguish two
different scenarios. 
Both are using a sharp measurement in the first step.
In the first scenario, the operator $A_i$ from
the first step is proportional to a unitary operator. 
This means that no information about the
input state is revealed, and we can choose that the
second step consists of a unitary operation described by only one
Kraus operator $B^{(i)}\sim A_i^\dagger$ \cite{nielsen97a}. If all
Kraus operators
fall into this scenario, we can in all events restore the input state
in the target qubit.

In the second scenario, the operator $A_i$ is not proportional to a
unitary operator, and
therefore the corresponding preliminary effective POVM element
$A_i^\dagger A_i \not\sim \openone$ reveals
some information about the input state. However, this can be
compensated, with some probability, by a measurement in the second
step. The idea is that one Kraus operator of the second step
`compensates' the knowledge, and we obtain for the overall effective
POVM element $A_i^\dagger
{B_j^{(i)}}^\dagger B_j^{(i)} A_i \sim \openone$. 

One can immediately determine the optimal choice of $B_j^{(i)}$, given
$A_i$, by looking at the singular value decomposition of $A_i= U
\left({{a}\atop{0}} {{0}\atop{b}}\right) V$ where $U,V$ are unitary
operators, and $1 \geq a \geq b \geq 0$ are constants. Then the choice
$B_j^{(i)} =V^\dagger \left({{b/a}\atop{0}} {{0}\atop{1}}\right)
U^\dagger$ results in $B_j^{(i)} A_i=b \openone$. This choice
guarantees the restoration of the input state in the target qubit.
Up to a pre-factor, the choice is unique. This 
pre-factor is constrained by the fact that 
the eigenvalues of $B_j^{(i)}$ may not exceed $1$, and 
in our choice
has been determined such that 
 it maximises the probability of  successful restoration.
Note that a restoration is not possible if $b=0$, corresponding to a
sharp effective measurement in the first step. This is in agreement
with Theorem \ref{sharpnesstheorem} which states that in this case the
target qubit is in a state independent of the input.

This formalism shows that any projection of the subsystems (except the
target qubit) onto a pure state allows a probabilistic restoration
unless this projection results in a sharp effective POVM
element. The possibility of a probabilistic restoration is therefore not uncommon. 
We will now illustrate the deterministic and the probabilistic restoration
in several settings.

\subsection{Deterministic restoration  in the ancilla or a clone}
\label{sec:resAnc}
In this scenario 
Alice has  coherent 
access to the two clones while Bob  has only access to the 
ancilla. They are also able to do one-way classical communication
from Alice to Bob. We now show
that it is possible to recreate the original input state in the
ancilla by a suitable measurement on the two clones, followed by a
conditional unitary dynamics on the ancilla. 

The output state of the cloner for an arbitrary input state is given
by
\begin{eqnarray}
U \left( \alpha \ket{0}+ \beta \ket{1} \right) \ket{0}
\ket{0}   
 & = &\
\sthird \ket{\phi^+} \left(- \beta \ket{0} + \alpha \ket{1}
\right)\nonumber \\
 &  &  +\sthird \ket{\phi^-}\left( \beta \ket{0} + \alpha \ket{1}
\right)\nonumber \\
 &  &+\sthird \ket{\psi^+}\left( - \alpha \ket{0} + \beta \ket{1}
\right)
\end{eqnarray}
Note that a measurement in the Bell basis on the two clones
corresponds to a measurement with effective POVM elements proportional
to the identity operator (following Eqn.~(\ref{EFformula})), 
thereby revealing no information about the
input state. Thus a Bell measurement on Alice's side allows the full
restoration of the input state. The measurement will
lead with equal probability of $1/3$ to one of the three symmetric Bell
states. Like in teleportation, Bob's remaining conditional state
 can then be transformed into the original input state 
by a suitable unitary operation, as shown in table \ref{ccteleport}.
\begin{table}
\[
\begin{array}[h]{l|r|c}
\mbox{result} & \mbox{conditional state}& \mbox{unitary operation}
\\\hline
\ket{\phi^+}&- \beta \ket{0} + \alpha \ket{1}& i \sigma_y
\\
\ket{\phi^-}&\beta \ket{0} + \alpha \ket{1}&\sigma_x
\\
\ket{\psi^+}&- \alpha \ket{0} + \beta \ket{1}& - \sigma_z
\end{array}\]
\caption{\label{ccteleport} A Bell measurement on the two
clones  leads to a conditional state of the
ancilla, which can be rotated unitarily into the original
input state. }
\end{table}

In a changed scenario where Alice has control of one clone and the
ancilla, and Bob has access to the other clone, Alice and Bob can
restore again the original input state in Bob's clone by means of a
Bell measurement of Alice and classical communication. To see this, we
 expand  the output  of the cloner for an arbitrary input
  in the Bell states of
the combined system clone-ancilla:
\begin{eqnarray}
\label{UcaBell}
U \left( \alpha \ket{0}+ \beta \ket{1} \right) \ket{0}
\ket{0}   
 & = &+\half \sqrt{3} \left(\alpha \ket{0}+ \beta \ket{1}\right) \ket{\psi^-}
 \nonumber \\
& & +\stwelvth\left(\alpha \ket{0}- \beta \ket{1}\right) \ket{\psi^+}\nonumber\\
& & + \stwelvth\left(\beta \ket{0}- \alpha \ket{1}\right) \ket{\phi^+}\nonumber\\
 &  & - \stwelvth\left(\beta \ket{0}+ \alpha \ket{1}\right)
\ket{\phi^-}\ .
\end{eqnarray}
 Alice performs a measurement in the Bell basis on her clone and the
ancilla. With probability $3/4$ she will obtain the result
 $\ket{\psi^-}$, and with probability $1/12$ each 
 one of the symmetric Bell states.  She communicates
the measurement result to Bob who performs the appropriate
transformation (see table \ref{acteleport}) to restore the input state
in his clone. 

\begin{table}
\[
\begin{array}[h]{l|r|c}
\mbox{result} & \mbox{conditional state}& \mbox{unitary operation}
\\\hline
\ket{\phi^+}& \beta \ket{0} - \alpha \ket{1}& - i \;\sigma_y
\\
\ket{\phi^-}&-\beta \ket{0} - \alpha \ket{1}&-\sigma_x
\\
\ket{\psi^+}& \alpha \ket{0} - \beta \ket{1}& \sigma_z
\\
\ket{\psi^-}& \alpha \ket{0} + \beta \ket{1}&\openone
\\
\end{array}\]
\caption{\label{acteleport} A Bell measurement on one of the
clones and the ancilla leads to a conditional state of the
other clone, which can be rotated unitarily into the original
input state.}
\end{table}

Therefore we have shown that it is not necessary to perform
the reverse cloning transformation
 on the total cloning output in order to recover
the original: the three-particle entangled output has the
property that measuring two subsystems and communicating classically
restores the original either in a clone or in the auxiliary
qubit. Thus  in  addition to the trivial 
way of restoration by revers cloning there is
another way which does not require to operate on
all output systems coherently. 

\subsection{Probabilistic restoration in the ancilla or a clone}

Let us now look at the case where the three output subsystems are
split up between three parties: Alice, Bob, and Charlie (who holds 
 the ancilla). No coherent measurement on two subsystems are possible now.
The parties    wish to restore
the original input state on Charlie's side by local measurements and
classical communication. We will show that this is possible,
although
with probability smaller than 1.

The cloning output for an arbitrary state can be written  as
\begin{eqnarray}
\label{UcBell}
U &&\left( \alpha \ket{0}+ \beta \ket{1} \right) \ket{0}
\ket{0} =  \nonumber \\
 &  &\ 
 \stwothird \alpha \ket{00} \ket{1} + \ssixth \ket{01} \left(- \alpha\ket{0}
  +\beta\ket{1}\right)\nonumber \\
 &  &-\stwothird \beta\ket{11}\ket{0} +\ssixth \ket{10}\left(- \alpha\ket{0}
  +\beta\ket{1}\right)\ .
\end{eqnarray}
Alice and Bob measure their clones in the  basis
$\{ 0,1 \}$. They
communicate their result to Charlie. If they find different outcomes,
 Charlie applies $-\sigma_z$, and the input
system is restored in his ancilla.
Otherwise, the corresponding
effective POVM element turns out to be sharp, 
so that the auxiliary system is in a
state independent of the input, and no recovery is possible.  However, the three parties know
 which situation occurred. The restoration is successful with
probability $1/3$.

We show now that it is also possible to restore the input state in a clone
held by Alice, while the other clone and the auxiliary system are held by
Bob and Charlie, respectively. We reorder the cloning transformation 
\begin{eqnarray}
\label{UpcBell}
U &&\left( \alpha \ket{0}+ \beta \ket{1} \right) \ket{0}
\ket{0} =  \nonumber \\
 &  & - \ssixth \alpha \ket{1}\ket{00}+\left( \stwothird \alpha
\ket{0}+\ssixth \beta \ket{1}\right) \ket{01}  \nonumber \\
& & + \left(- \ssixth \alpha \ket{0}-\stwothird \beta \ket{1}\right)
\ket{10}+\ssixth \beta \ket{0}\ket{11} \; .
\end{eqnarray}
Bob and Charlie measure their system in the 
 basis
$\{ 0,1 \}$ and
communicate their result to Alice. If the two outcomes are different,
then Alice can restore the original input with some probability by
application of a filter operation. 
If Bob has found ``0'' and Charlie has found
``1'', Alices successful 
 filter operation
is described by the Kraus operator $A_F = \frac{1}{2}
\ket{0}\bra{0}+\ket{1}\bra{1}$ . If the results of Bob and Charlie are 
interchanged, Alice applies
a filtering operation such that the success is described by
$A_F =-\ket{0}\bra{0}-\frac{1}{2}\ket{1}\bra{1}$. The total
probability of success, including the measurements of Bob and Charlie
and the filtering operation, is again $1/3$. 

Note that our restoration scheme 
is related to ideas of quantum secret sharing
in the sense of \cite{hillery99a}: only when the parties are
cooperating they are able to fully restore the original. Each of them
does have some knowledge about the state, though, and therefore the
cloning output state does not correspond to secret sharing in the
spirit of \cite{cleve99a}, where none of the subsystems contains any
information.

\section{Summary and Discussion}
In this paper we have studied various measurements on parts of the
output of a $1\to 2$ universal cloning machine. We have shown their
correspondence to effective measurements on the input qubit.  

In particular, we were interested in sharp effective measurements, as
they are the ones that maximise the information gain and fidelity of
state estimation, and allow state exclusion.
 We have found
that measurements on any two output subsystems can correspond to sharp
effective measurements at the input, whereas measurements on one
output only can not. A complete set of sharp effective measurements at
the input, though, can only be implemented by measurements on the two
clones.

 We have also studied the possibility of restoring the
original in one of the output qubits, after performing certain
measurements on the other outputs, and communicating classically. In
scenarios where a party holds two of the three outputs this
teleportation-like scheme was shown to operate with probability
one. In a scenario where each party holds only one output the
restoration is successful with probability smaller than one. 

 We
hope that these results may be a step towards answering the question
whether approximate quantum cloning is a useful process in quantum
information processing.

\section{Acknowledgements}
We gratefully acknowledge helpful discussions with Maciej
Lewenstein, Kalle-Antti Suominen and Guifre Vidal. 
The authors 
 took benefit from the Workshop on Quantum Information Processing, 
       Benasque Center for Science, 2000.
       We acknowledge support by   the ESF PESC Programme
       on Quantun Information, and by the IST Programme EQUIP.
DB has been supported by 
Deutsche Forschungsgemeinschaft under 
SFB 407 and Schwerpunkt  QIV.
JC and NL have been funded by the Academy of Finland under project
43336 and by the Academia Scientiarum Fennica.


\end{document}